\documentstyle[12pt]{article}
\begin{document}
\title{Pedagogical Introduction to Hamiltonian BRST formalism}
\author{NAOHISA OGAWA \\
Hokkaido University of Education Physics Department,\\
Sapporo 002, JAPAN.}
\maketitle
\begin{abstract}
Hamiltonian BRST formalism (FV formalism) includes many auxiliary fields without 
explanation. Its path-integration has a simple form by using BRST charge, but its 
construction is quite mechanically and hard to understand physical meaning.  
In this paper we perform the phase space path-integral with requiring BRST invariance 
for action and measure, and show that the resultant form is equivalent to  
the Hamiltonian BRST (FV) formalism in gravitational theory. 
This explains why so many auxiliary fields are necessary to be introduced.  
We also find the gauge fixing is automatically done by requiring the BRST  
invariance of the path-integral measure.  
This is a pedagogical introduction to Hamiltonian BRST formalism. 
\end{abstract}
\

  We consider the gravitational theory at D+1 dimension in phase-space action.
 The reason why we should start from such an action is that the path-integral measure can not be defined in configuration space when its symmetry is non compact such as, the case of time reparametrization invariant systems including the gravity. Our notation is as follows: the small Latin indices: $\{i,j,k, \ldots \}$ run from 1 to D (space-dimension), and the Greek indices: $\{ \alpha,\beta, \mu, \ldots \}$ run from 0 to D.  We use the ADM decomposition for metric tensor: $ g_{ij} $ as spacial metric, and $ N^{\mu}$ as laps-function and shift-vector.  Another dynamical variable is the canonical momentum: $\pi^{ij}$ which is conjugate to $g_{ij}$.
Raising and lowering the Latin indices are performed by using the spacial metric: $g_{ij}$ and $g^{ij}$.  By using the above variables, our Einstein-Hilbert action can be written in the form;
\begin{equation}
S_{grav.} \: \equiv\: \int dt \, L^{grav.}(t) \: \equiv \: \int dt \:[\; \pi^{ij} \dot{g}_{ij}\:-\:N^{\mu}T_{\mu}\;].
\end{equation}
The contraction of indices is defined to include the spacial D-dimensional integration.  $ \{ T_{\mu} \} $ are the first class constraints which satisfy the poisson bracket's relation,
\begin{equation}
       \{ T_{\mu}, \: T_{\nu} \}\: = \: U^{\lambda}_{\mu\nu} T_{\lambda}.
\end{equation}
The form of these constraints are as well known,
\begin{equation}
         T_0 \:\equiv \: \frac{1}{\sqrt{g}}[\,g R + \pi^{ij} \pi_{ij} - \frac{1}{2} \pi^2 \,], \:\:\: T_i \:\equiv \: -2 \nabla_j \, \pi^j_i.
\end{equation}
 Here we should note that $ \pi^{ij} $ is the tensor density, and $ \pi \equiv g_{ij} \pi^{ij} $. The Poisson bracket is defined on phase space $(g_{ij}(x), \pi^{ij}(x))$. One of the structure functions is depending on spacial metric field as,
\begin{equation}
  U^{i(z)}_{0(x),0(y)} = \delta(x-z) g^{ik}(x) \partial^x_k \delta(x-y) -  
  (x \leftrightarrow y),
\end{equation}
and others are independent of the fields.  The gauge-transformation is defined by using the 1-st class constraints as  \cite{FV},
\begin{equation}
 \delta \pmatrix{ g_{ij} \cr
 \pi^{ij}\cr}
  = \{ \pmatrix{ g_{ij} \cr
 \pi^{ij}\cr}
, T_{\mu} \} F^{\mu} ,\:\: 
\delta N^{\mu} = \dot{F}^{\mu} - U^{\mu}_{\beta \gamma} N^{\beta}F^{\gamma}.
\end{equation}
The explicit calculation shows:
\begin{equation}
   \delta L^{grav.} = \frac{d}{dt} [\,  F^{\mu} \, (\pi^{ij} \{ g_{ij}, T_{\mu} \} - T_{\mu}) \,].
\end{equation}
Then we have the invariance of canonical action with the boundary condition: 
$ F^{\mu}(t_{initial}) = F^{\mu}(t_{final}) = 0 $.  Let us consider the path-integral measure firstly.  The Jacobian under the above gauge transformation for each field can be calculated explicitly, and we find 
\begin{equation}
                 \prod_{i \leq j,x,t}\: dg_{ij}(x,t) \, d\pi^{ij}(x,t),
\end{equation}
is invariant under the gauge transformation.
This is nothing but Liouville theorem. Using $\phi_n(x)$, which is the orthogonal complete set in real functional space satisfying,
\[\int d^Dx \; \phi_n(x) \phi_m(x) = \delta_{mn},\:\:\: \sum_n \phi_n(x) \phi_n(y) = \delta^D(x-y),\]
 we obtain the Jacobian for $N^{\mu}$ as:
\begin{equation}
 J(N^{\mu}) = \exp[-\frac{D+3}{2} \sum_n \int d^Dx \, \phi^2_n(x) \, \partial_k F^k].
\end{equation}
Since $N$ is not the canonical variable, this variable does not have a partner for its Jacobian to be canceled with. The existence of this unpleasant Jacobian factor is essential for gravity, since this measure is invariant in Yang-Mills case and also in the relativistic free particle case. We find that Fujikawa's technic \cite{FUJI} which modifies the variable to hold the gauge invariance of the measure does not help us from this difficulty. Since we can not construct the invariant measure for basic fields, Faddeev-Popov method \cite{FP} is hardly possible, and we should work with Fujikawa's method \cite{FUJI} (by Fujikawa's method, I mean the requirement of BRST invariance of action and path-integral measure: anomaly free condition).  The another problem appears at nilpotency for BRST transformation. We define the BRST transformation by rewriting $F^{\mu}$  to ghost field: $C^{\mu}$ as,
\begin{equation}
\delta^B g_{ij}= \{g_{ij}, T_{\mu} \} C^{\mu},\: \delta^B \pi^{ij}= \{\pi^{ij}, T_{\mu} \} C^{\mu},\:  \delta^B N^{\mu} = \dot{C}^{\mu} - U^{\mu}_{\alpha \beta} N^{\alpha} C^{\beta}.
\end{equation}
The BRST transformation of the ghost field is determined by the nilpotency condition of fields. Requiring the nilpotency condition on the metric field, and by using the Jacobi-identity for $g_{ij}$, $T_{\mu}$, and $T_{\nu}$, we obtain
\begin{equation}
  \delta^B C^{\lambda} = \frac{1}{2}\, U^{\lambda}_{\mu\nu} C^{\mu}C^{\nu},
\end{equation}
where, we should remember that each contraction of indices includes spacial integration. This relation is quite similar as the non-abelian gauge theory except the difference between structure-constant and structure-function.
\

    The nilpotency condition for ghost field:
\begin{equation}
(\delta^B)^2 \,C^{\mu} \:= \:0.
\end{equation}
is guaranteed by the generalized Jacobi-identity:
\begin{eqnarray}
 &{}& \{U^{\lambda}_{\mu\nu}, T_{\rho}\} + \{U^{\lambda}_{\nu\rho}, T_{\mu}\} + \{U^{\lambda}_{\rho\mu}, T_{\nu}\}\nonumber \\
&{}& \: \: + U^{\sigma}_{\mu\nu}U^{\lambda}_{\sigma\rho}\: + U^{\sigma}_{\nu\rho}U^{\lambda}_{\sigma\mu}\: + U^{\sigma}_{\rho\mu}U^{\lambda}_{\sigma\nu}\: = \:0.
\end{eqnarray}
But for the momentum field, we obtain
\begin{equation}
  (\delta^B)^2 \, \pi^{ij}(x) = \frac{1}{2} \, \frac{\delta U^{\lambda}_{\mu\nu}}{\delta g_{ij}(x)}\, T_{\lambda} \, C^{\mu} \, C^{\nu}.
\end{equation}
And also for the multiplier field,
\begin{equation}
(\delta^B)^2 \, N^{\lambda}(x) \:=\: \frac{1}{2} \, \frac{\delta U^{\lambda(x)}_{\mu\nu}}{\delta g_{ij}}\,(\dot{g}_{ij}- \frac{\delta H}{\delta \pi^{ij}})\, C^{\mu} \, C^{\nu},
\end{equation}
where $H$ is the classical total Hamiltonian: $N^{\mu}T_{\mu}$.  In this way the nilpotency is broken on $\pi^{ij}$ and $N^{\mu}$ fields, because of the field dependence of structure function.

   We have two problems now, that is, (I) invariant measure for multiplier field, (II) nilpotency of BRST transformation. Firstly we start with the invariant measure problem. Since the modification of variable $N^{\mu}$ could not solve this problem, we need to introduce some other method which perfectly insures the gauge invariance. The best way ever we know is to introduce the canonical partner of $N$ field, and to depend on the Liouville theorem. Introducing the new field $P_{\mu}$, and require the invariance of measure: $\prod \, dN^{\mu}\,dP_{\mu}$.
If we use the Liouville theorem to hold the gauge invariance of the measure, it is necessary that $P$ and $N$ are canonically conjugate each other. 
Therefore the Lagrangian should contain the kinetic term: $ P_{\mu}\,\dot{N}^{\mu}$. Then $N$ is not the multiplier field but dynamical one, even though it is not physical field. So we should include it in a member of BRST quartet  \cite{KUGO} to insure its non-physicality at the last stage. Next this change of the Lagrangian breaks its local gauge invariance, and so we have to go to BRST symmetry quite naturally without any gauge fixing by hand. We must find such a BRST symmetry which should satisfy three conditions. Firstly it contains $P$ and $N$ as the member of BRST quartet, and that should insure the invariance of their measure. Secondary it should be the symmetry of the total Lagrangian. Thirdly it should be nilpotent on all the fields. We should remark that this situation is much different from the other gauge theories, where we can decide the gauge condition by hand without considering the measure's invariance. In Fradkin-Vilkovsky formalism \cite{FV}, the gauge condition: $\dot{N}^{\mu}+$(something)~ was utilized to obtain the explicit Lorentz covariant gauge, but here such a gauge was introduced to insure the invariance of the path-integral measure automatically. In the above sense, we should firstly define the BRST transformation, and next check the above consistency conditions.
\

    Now we come to the nilpotency-problem. Let us begin with introducing the BRST charge which generates the transformation: $\delta^B g_{ij}$ and $\delta^B C^{\mu}$ (and so the nilpotency on $g_{ij}$ also,) which is obtained earlier, in the form,
\newcommand{\brs}{\tilde{\delta}^B}
\begin{equation}
     \brs A(x)\: \equiv \: - \, \{ \, Q_{B}, \: A(x) \, \}.
\end{equation}
To introduce the transformation of the ghost, we have extended the Poisson bracket as,
\begin{equation}
 \{\, A, \: B \, \} \equiv \sum_a [\, \partial A / \partial q^a \cdot \frac{\partial}{\partial p_a}B\: - \: (-)^{\mid a \mid} \partial A / \partial p_a \cdot \frac{\partial}{\partial q^a} B \,],
\end{equation}
where $\mid a \mid$ is 1 for fermionic field "a", and 0 for bosonic field "a", and, $ q^a = (g_{ij}, C^{\mu}), \:\:\: p_a = (\pi_{ij}, \bar{C}_{\mu})$.
The form of the BRST charge which satisfies the required condition is easily 
found as,
\begin{equation}
     Q_{B} = T_{\mu} C^{\mu} - \frac{1}{2} \bar{C}_{\lambda} U^{\lambda}_{\mu\nu}\, C^{\mu} C^{\nu}.
\end{equation}
But we find that this transformation is different from the previous one on momentum field as,
\begin{equation}
  \brs \pi^{ij} \equiv -\{Q_B, \pi^{ij} \} = \{\pi^{ij}, T_{\mu} \}C^{\mu} + \frac{1}{2} \bar{C}_{\lambda}\frac{\delta U^{\lambda}_{\mu\nu}}{\delta g_{ij}}\, C^{\mu} C^{\nu}.
\end{equation}

However, we can prove the nilpotency for this BRST charge by using the Jacobi-identity and the anticommuting nature of the ghost field,
\begin{equation}
 \{ Q_B, \: Q_B \}\, = \, 0.
\end{equation}
And also for any field $A(x)$, by using the Jacobi-identity we can prove that
\begin{equation}
 (\brs)^2 A(x) = -\frac{1}{2} \{A(x), \{Q_B, Q_B\}\} = 0.
\end{equation}
Therefore  the BRST transformation $\brs$ is completely nilpotent on any fields, and so is more preferable than to use $\delta^B$. 
But we notice that since $\brs$ is different from the original gauge transformation, the original Lagrangian is no longer BRST invariant. This problem is solved by considering the BRST transformation of $N$ field in the following. The BRST transformation for $N$ was originally defined by (9). But now we would like to rewrite it in the form (15). The best way to do it is that, we define the BRST transformation of $N$ as the original BRST transformation for $N$ is reproduced on on-shell condition as follows.
Supposing that the equation of motion:
\begin{equation}
   \tilde{C}^{\mu} = \dot{C}^{\mu} - U^{\mu}_{\alpha\beta} N^{\alpha} C^{\beta},\end{equation}
holds as Euler-Lagrange equation, we redefine the total BRST charge as
\begin{equation}
  Q_B \: \equiv \: T_{\mu} C^{\mu} - \frac{1}{2} \bar{C}_{\lambda} U^{\lambda}_{\mu\nu}\, C^{\mu} C^{\nu} + P_{\mu}\tilde{C}^{\mu}.
\end{equation}
Here we introduced the new ghost field $\tilde{C}^{\mu}$, and we also introduce its conjugate field $\bar{\tilde{C}}_{\mu}$ implicitly. All the canonical variables are now
\begin{equation}
  q^a :\:(g_{ij},\, C^{\mu},\, N^{\mu},\, \tilde{C}^{\mu}),\:\:
  p_a :\:(\pi^{ij},\, \bar{C}_{\mu},\, P_{\mu},\, \bar{\tilde{C}}_{\mu}),
\end{equation}
and our Poisson bracket is also extended by the above variables.
Note that the new BRST charge stays still to be nilpotent. Let us summarize our BRST transformation.
\begin{eqnarray}
&\brs g_{ij}& = \frac{\delta T_{\mu}}{\delta \pi^{ij}} \cdot C^{\mu},\:\:
\brs \pi^{ij} = \, - \frac{\delta T_{\mu}}{\delta g_{ij}} \cdot C^{\mu}\,+ \,\frac{1}{2} \frac{\delta U^{\lambda}_{\mu\nu}}{\delta g_{ij}} \, \bar{C}_{\lambda}C^{\mu}C^{\nu},\\
&\brs C^{\mu}& = \frac{1}{2} U^{\mu}_{\alpha \beta} C^{\alpha}C^{\beta},\:\:
\brs \bar{C}_{\mu} = -(T_{\mu} + U^{\lambda}_{\mu\nu}\bar{C}_{\lambda}C^{\nu}),\\
&\brs N^{\mu}& = \tilde{C}^{\mu}, \:\: \brs P_{\mu} = 0, \:\: \brs \tilde{C}^{\mu} = 0, \:\: \brs \bar{\tilde{C}}_{\mu} = - P_{\mu}.
\end{eqnarray}
We should remark three points here. Firstly since the modified part of the BRST transformation for $\pi^{ij}$ does not contain $\pi^{ij}$ itself, this modification does not change the invariance of the path-integral measure: $dg d\pi$. Secondary not only $(N^{\mu}, P_{\mu})$ but also other sets of ghosts are forming the BRST invariant measure due to the Liouville theorem. Therefore the path-integral measure is BRST invariant. Thirdly the last equations show that the set: $(N,\tilde{C},\bar{\tilde{C}},P)$ is forming the BRST quartet. Then we do not have to mind on the dynamical degree of $N$ and $P$ fields as well as the related ghost fields.
\

   Now the remained problem is to find the BRST invariant action which includes equation (21), and it should include the kinetic part of additional ghost pair $(\tilde{C},\bar{\tilde{C}})$ to hold the quartet mechanism. The simplest solution for this problem is easily found as,
\begin{equation}
 L = \pi^{ij}\dot{g}_{ij} - N^{\mu}T_{\mu} + P_{\mu}\dot{N}^{\mu} + \bar{\tilde{C}}_{\mu}\dot{\tilde{C}}^{\mu} + \bar{C}_{\mu} \{\dot{C}^{\mu}-U^{\mu}_{\alpha\beta}N^{\alpha}C^{\beta}-\tilde{C}^{\mu} \}.
\end{equation}
The variation by $\bar{C}_{\mu}$ gives the equation (21), and the variation by $P_{\mu}$ gives the gauge condition $\dot{N}=0$. Actually the local gauge freedom is completely fixed by this term since other terms can not generate the term: $P_{\mu}\ddot{F}^{\mu}$ under the local gauge transformation to cancel with. The BRST invariance of this Lagrangian is simply found by rewriting it in the form;
\begin{equation}
L = \pi^{ij}\dot{g}_{ij} + P_{\mu}\dot{N}^{\mu} + \bar{\tilde{C}}_{\mu}\dot{\tilde{C}}^{\mu} + \bar{C}_{\mu}\dot{C}^{\mu} - \{\, Q_B, \,N^{\mu}\bar{C}_{\mu} \}.\end{equation}
Since the canonical transformation does not change the form: $p_a \dot{q}^a$, each kinetic term is BRST invariant, and the nilpotency insures the invariance of the remained term. Therefore we could construct the BRST invariant measure and action only by using the canonical-action and canonical-symmetry. This is our goal for Einstein gravity, and is really the same as the ones given by Fradkin-Vilkovsky with special gauge (gauge-function: $\chi(x,\pi,N,P) = 0$) \cite{FV}.
Once we have obtained the BRST invariant action, it is easy to generalize the gauge condition.  We take the gauge function: $\chi$, and change the gauge condition as,
\begin{equation}
 \dot{N}^{\mu} = 0 \:\: \longrightarrow \:\: \dot{N}^{\mu}-\chi^{\mu}(g,\pi,N) = 0.
\end{equation}
This can be done in BRST invariant Lagrangian smoothly as,
\begin{eqnarray}
L^{gen.} &=& \pi^{ij}\dot{g}_{ij} + P_{\mu}\dot{N}^{\mu} + \bar{\tilde{C}}_{\mu}\dot{\tilde{C}}^{\mu} + \bar{C}_{\mu}\dot{C}^{\mu} - H_{gen.},\\
H_{gen.} &=& \{\, Q_B, \,N^{\mu}\bar{C}_{\mu} + \chi^{\mu}\bar{\tilde{C}}_{\mu} \}.
\end{eqnarray}
The explicit form of the general Hamiltonian is as follows,
\begin{eqnarray}
H_{gen.} &=&  N^{\mu}T_{\mu} + P_{\mu}\chi^{\mu} + \bar{C}_{\mu}U^{\mu}_{\alpha\beta}N^{\alpha}C^{\beta} + \bar{C}_{\mu}\tilde{C}^{\mu}\nonumber \\
&{}& + \bar{\tilde{C}}_{\mu} \{ \chi^{\mu}, T_{\nu} \} C^{\nu} + \bar{\tilde{C}}_{\mu}\frac{\delta \chi^{\mu}}{\delta N^{\nu}} \tilde{C}^{\nu} + \frac{1}{2} \bar{\tilde{C}}_{\alpha}C^{\beta} \{ \chi^{\alpha}, \, U^{\rho}_{\beta\mu} \} \bar{C}_{\rho} C^{\mu},
\end{eqnarray}
which is  the same as Fradkin, Vilkovsky's one in general gauge.
 It is also possible to extend the functional form of $\chi$ as to depend on $P_{\mu}$. Then the system is still BRST invariant and this extension of gauge function admits us to change from delta function type gauge fixing to Fermi type gauge fixing in the path-integral.

   We summarize our results here. We took a quantization rule to preserve the canonical-symmetry in quantum level. Use of this rule for the gravity gave the same result as Fradkin-Vilkovsky's one. In their previous work, the path-integral form as solution was proved to be meaningful in theorem, and the method to obtain its form was not clear. In this paper we have shown the simple derivation of Fradkin-Vilkovsky's path-integral form, and showed that the gauge fixing was automatically done by the BRST invariance of the measure, which is most different point from other gauge theories. \newline
The author would like to thank Prof.M.Henneaux for reading through the manuscript and 
helpful comment.  He also thanks Prof.K.Fujii for encouraging him.
This work is partly supported by Deutscher Akademischer Austauschdienst.

\end{document}